\documentclass[aps,superscriptaddress,prl,twocolumn,showpacs,letterpaper,tighten,float]{revtex4-1}
\usepackage{amssymb}
\usepackage{amsbsy}
\usepackage{amsmath}
\usepackage{epsfig}
\usepackage{graphicx}
\usepackage{array}
\usepackage[utf8]{inputenc}
\usepackage{textcomp}
\usepackage{color}
\usepackage{bm}

\definecolor{myblue}{rgb}{.93, .93, 1}
\newcommand*\mybluebox[1]{%
\colorbox{myblue}{\hspace{1em}#1\hspace{1em}}}
\newcommand*\myFbluebox[1]{%
\fcolorbox{black}{myblue}{\hspace{1em}#1\hspace{1em}}}

\newcommand{\bsub}{\begin{subequations}}
\newcommand{\esub}{\end{subequations}}
\newcommand{\beq}{\begin{empheq}[box=\mybluebox]{align}}
\newcommand{\beqF}{\begin{empheq}[box=\myFbluebox]{align}}

		\newcommand{\ord}[1]{\bm{\mathit{O}}\left(#1\right)}

\begin{document}

\title{Helical Quantum Edge Gears in 2D Topological Insulators}
\author{Yang-Zhi~Chou} \email{yc26@rice.edu}
\affiliation{Department of Physics and Astronomy, Rice University, Houston, Texas 77005, USA}
\author{Alex~Levchenko}
\affiliation{Department of Physics, University of Wisconsin-Madison, Madison, Wisconsin 53706, USA}
\author{Matthew~S.~Foster}
\affiliation{Department of Physics and Astronomy, Rice University, Houston, Texas 77005, USA}
\affiliation{Rice Center for Quantum Materials, Rice University, Houston, Texas 77005, USA}

\date{\today}

\pacs{71.10.Pm, 72.15.Nj, 74.25.F-}

\begin{abstract}
We show that two-terminal transport can measure the Luttinger liquid (LL) parameter $K$, in helical LLs  at the edges of two-dimensional topological insulators (TIs) with Rashba spin-orbit coupling. We consider a Coulomb drag geometry with two coplanar TIs and short-ranged spin-flip interedge scattering. Current injected into one edge loop induces circulation in the second, which floats without leads. In the low-temperature ($T \rightarrow 0$) perfect drag regime, the conductance is $(e^2/h)(2 K + 1)/(K + 1)$.  At higher $T$ we predict a conductivity $\sim T^{-4K+3}$. The conductivity for a single edge is also computed. 
\end{abstract}

\maketitle

The edge states that encircle two-dimensional (2D) topological insulators (TIs)
realize a novel electronic helical Luttinger liquid (HLL) phase \cite{Kane2005_1,Hasan2010_RMP,Qi2011_RMP}. Distinct from an ordinary one-dimensional (1D)
quantum wire and from a quantum Hall edge, a helical edge consists of two counterpropagating 
modes forming a Kramers pair. The left- and right-moving channels interact through Coulomb repulsion, 
but time reversal symmetry protects the edge from the opening of a gap and from Anderson localization due to impurities. The combination of topological protection and electron correlations implies that a 
TI edge is an ideal Luttinger liquid at low temperatures \cite{Xu2006,Wu2006}. Experimental evidence for helical edge states in HgTe \cite{Konig2007} and InAs/GaSb \cite{Knez2011} includes a quantized conductance $G \simeq 2 e^2 / h$ \cite{Roth2009,Knez2011}. 

In the absence of electrical contacts and magnetic fields, a HLL forms a closed, unbreakable loop. This \emph{topology of the edge} has so far received little attention. In this Letter, we propose a TI device geometry in which edge loops rotate as interlocking ``gears'' through Coulomb drag \cite{Rojo1999,Nazarov1998,Ponomarenko2000,Klesse2000,Narozhny2015}. 
Our main result is that the strength of electron correlations encoded in the Luttinger parameter 
can be directly obtained in such a device using a two-terminal dc conductance measurement. 

Correlations are generically strong in 1D electron fluids because two particles cannot exchange positions without scattering or tunneling. These correlations are encoded in the Luttinger parameter $K$ \cite{Giamarchi_Book}. Measuring $K$ in a nontopological 1D electronic system (or ``wire'') is 
possible but delicate. For instance, the zero-temperature ($T = 0$) dc transport through a perfectly clean wire gives a quantized conductance independent of $K$ \cite{Maslov1995,Ponomarenko1995,Safi1995}. In a long wire, disorder tends to induce Anderson insulating behavior. At temperatures $T \gtrsim \hbar v k_F / k_B$, inelastic scattering due to irrelevant umklapp interactions gives a conductivity that depends on $T$ through a power law \cite{Giamarchi91}; here $v$ and $k_F$, respectively, denote the charge velocity and Fermi wave vector. The disorder-induced scattering may lead to a qualitatively similar effect \cite{Maslov}. The temperature exponent in conductance can reveal the Luttinger parameter $K$, but a large temperature range is needed to fit the data. The tunneling zero bias anomaly is also predicted to encode $K$, but measurements often contain contributions from other mechanisms \cite{Deshpande2010}.

\begin{figure}[b]
\includegraphics[width=0.33\textwidth]{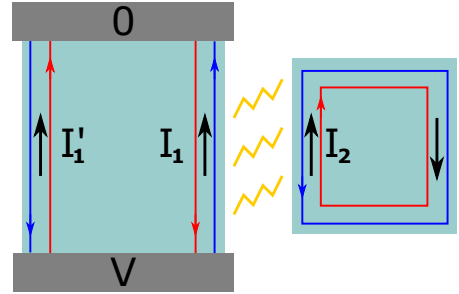}
\caption{Using helical quantum edge gears to measure the Luttinger parameter. We consider $\mathbb{Z}_2$ TI edge states in two adjacent topological regions. The blue and red arrows indicate the propagation directions of edge electrons with opposite helicities. The left TI is connected to external leads; $I_1$ and $I_1'$ denote the currents of the edges connected to these. The right TI edge floats as an electrically isolated closed loop. Rashba spin-orbit coupling \cite{Kane2005_1} enables Coulomb drag due to short-ranged spin-flip scattering  \cite{Tanaka2009} between the adjacent edges.
This induces a current $I_2$ that circulates in the right edge. In the case of identical TIs with an interacting edge region of size $L \rightarrow \infty$, at zero temperature strong backscattering ``locks'' the currents $I_1=I_2$, associated to perfect drag \cite{Nazarov1998,Klesse2000}. We then predict that the zero-temperature conductance is  $G=(I_1+I_1')/V=(e^2/h)(2 K + 1)/(K + 1)$, where $K$ is the Luttinger parameter. In a real system of finite length $L \gg \xi$ and at temperatures $T$ satisfying $\hbar v / L  \lesssim k_B T \ll \Delta$ \cite{Ponomarenko2000} with $\xi = \hbar v / \Delta$ and $\Delta$ 
the Mott gap of the antisymmetric mode, the result for $G$ holds up to terms exponentially small in 
$L / \xi$ and $\Delta / k_B T$ \cite{Nazarov1998,Ponomarenko2000,Klesse2000}.  
Here $v$ is the charge velocity.}
\label{Fig:2Edge_scheme}
\end{figure}

In the simplest version of HLL physics that realizes the quantum spin Hall effect \cite{Kane2005_2,Bernevig2006,Xu2006,Wu2006}, the $z$ component of spin is 
assumed to be conserved in a TI. As a result, the edge electrons carry well-defined $S_z$ currents. 
When Rashba spin-orbit coupling (SOC) is present \cite{Kane2005_1} (generically expected in the absence of inversion symmetry), $S_z$ symmetry is sabotaged. New spin-flip interactions \cite{Tanaka2009,Schmidt2012,Lezmy2012} are then allowed on TI edges. 

We show that the Luttinger parameter enters the conductance in a Coulomb drag geometry consisting of two coplanar TI regions with Rashba SOC. Over a segment of length $L$, proximate HLL edge states are separated by a gap narrow enough to allow short-ranged Coulomb scattering but wide enough to prevent tunneling. In Fig.~\ref{Fig:2Edge_scheme}, we consider two identical helical edges. Current $I_1$ is injected by external leads. Short-ranged spin-flip scattering \cite{Tanaka2009} between edges induces a current $I_2$ in the right TI edge loop, which floats without leads. At zero temperature, the two proximate edge segments develop a locking state of perfect drag ($I_1 = I_2$) \cite{Nazarov1998,Klesse2000} 
for an infinitely long interacting region $L \rightarrow \infty$. An additional current $I_1'$ flows in parallel between the contacts. The zero-temperature two-terminal conductance $G = (I_1 + I_1')/V$ is
\begin{align}\label{GResult}
	G 
	= 
	\frac{e^2}{h}\left[1 + \left(1 + 1/K\right)^{-1}\right] 
	= 
	\frac{e^2}{h} \left(\frac{2 K + 1}{K + 1}\right),
\end{align}
where $(1 + 1/K)$ is the dimensionless resistance of the locked edges, as explained below.
For a finite locking length $L \gg \xi$ and at temperatures $T$ satisfying $\hbar v / L  \lesssim k_B T \ll \Delta$ \cite{Ponomarenko2000}, Eq.~(\ref{GResult}) holds up to exponentially small corrections in 
$L / \xi$ and $\Delta / k_B T$ \cite{Nazarov1998,Ponomarenko2000,Klesse2000}. Here $\xi \equiv \hbar v / \Delta$ is the length scale associated to the gapped ``antilocking'' mode with $I_1 = - I_2$; $\Delta$ is the energy gap. 

We also discuss dissipative finite-temperature transport in this geometry. In contrast to the usual setup for Coulomb drag \cite{Rojo1999,Narozhny2015}, the system is naturally characterized in terms of conductances or conductivities: 
\begin{align*}
\left[\begin{array}{c}
I_1\\
I_2
\end{array}
\right]\hspace{-0.75mm}=\hspace{-0.75mm}
\left[
\begin{array}{cc}
G_{11} & G_{12}\\
G_{21} & G_{22}
\end{array}
\right]
\left[\begin{array}{c}
V_1\\
V_2
\end{array}
\right],
\;\;\;
G_{i j} = \sigma_{i j} / L,
\end{align*}
where the labels $1$ and $2$ indicate the active and passive systems respectively. For our TI edges, the passive system is a closed HLL loop with $V_2 = 0$, $I_1=\sigma_{11} V_1 / L$, and $I_2=\sigma_{21} V_1 / L$. We compute the intraedge and transconductivities using the Kubo formula and bosonization,
employing the effective potential formalism \cite{Peskin_Book,Ristivojevic12}. Both $\sigma_{11}$ and $\sigma_{21}$ give $T^{-4K+3}$ ($T^{-4 K + 2}$) behavior in the absence (presence) of disorder, above the locking transition.

Finally, we compute the conductivity of a single edge due to the least irrelevant symmetry-allowed 
(one-particle umklapp) interaction term. We find asymptotic $T^{-2K-1}$ ($T^{-2K-2}$) behavior in the high- (low-) $T$ limits, in the presence of disorder, consistent with \cite{Kainaris2014}, and we also obtain the full result for the clean limit. 
Power-law scaling of conductance as a function of temperature and bias voltage 
that may be attributable to Luttinger liquid physics
was recently observed in InAs/GaSb quantum spin Hall devices \cite{RRD-arXiv}.


{\it Model}. -- The edge states of a 2D TI can be expressed in terms of right ($R$) and left ($L$) mover fermion fields. The kinetic term is 
\begin{align}\label{Free_F}
	\hat{H}_0
	=
	-i \hbar v_F
	\int dx
	\left[
	R^{\dagger}(x) 
	\partial_x 
	R(x)
	-
	L^{\dagger}(x) 
	\partial_x
	L(x)
	\right],
\end{align}
where $v_F$ is the Fermi velocity of the edge band. Time-reversal symmetry is encoded by $R(x)\rightarrow L(x)$, $L(x)\rightarrow -R(x)$, and $i\rightarrow -i$. Left and right movers interact via 
intraedge Coulomb repulsion.

We focus on the coplanar geometry in Fig.~\ref{Fig:2Edge_scheme} and consider the backscattering components of the interedge Coulomb interaction. An additional interedge Luttinger interaction does not modify our results for the locking regime if the distal portion of edge loop 2 is much longer than the interacting segment of length $L$; otherwise the parameter $K$ in Eq.~(\ref{GResult}) encodes a combination of inter- and intraedge correlations. In the presence of Rashba SOC, the following interedge backscattering terms are allowed by symmetry \cite{Tanaka2009}: 
\begin{align}
	\label{H_I_-}
	\hat{H}_{-}=&\, U_-\int dx \left[e^{i2(k_{F1}-k_{F2})x}L^{\dagger}_1R_1R^{\dagger}_2L_2	+ \text{H.c.}	\right],
	\\
	\label{H_I_+}
	\hat{H}_{+}=&\, U_+\int dx \left[e^{i2(k_{F1}+k_{F2})x}L^{\dagger}_1R_1L^{\dagger}_2R_2	+ \text{H.c.}	\right],
\end{align}
where $k_{F1}$ ($k_{F2}$) indicates the Fermi momentum in the first (second) edge. These are defined relative to an edge Dirac point, which is a commensurate (time-reversal invariant) momentum \cite{Hasan2010_RMP}. The $U_-$ interaction describes normal backscattering, while $U_+$ is a two-particle umklapp interaction. Additional one-particle umklapp interaction terms are also allowed, 
\begin{align}
	\nonumber
	\hat{H}_{U}
	=
	\sum_{a=1,2}
	U_{a}
	\int dx \,&
		\Big[	e^{-i2k_{Fa}\,x}	R^{\dagger}_a L_a R^{\dagger}_{\bar{a}} R_{\bar{a}}\\
	\label{H_U}
			&-e^{i2k_{Fa}\,x}	L^{\dagger}_a R_a L^{\dagger}_{\bar{a}} L_{\bar{a}}+\text{H.c.}
		\Big],
\end{align}
where $a$ is the index of the edge, $\bar{1}=2$, and $\bar{2}=1$. It is worth mentioning that 
all of these interactions are disallowed in the presence of $S_z$ conservation (in each edge) 
\cite{Tanaka2009,Zyuzin10}. 

For simplicity, we assume the two HLLs are identical, so that $k_{F1} = k_{F2} \equiv k_F$ and $U_1 = U_2 \equiv U$. The dominant interedge interaction at $T=0$ is the nonumklapp backscattering $\hat{H}_-$; the others are irrelevant at long wavelengths for $k_F \neq 0$ \cite{Giamarchi_Book}. In order to include Luttinger liquid effects, we use bosonization \cite{Shankar1995,Giamarchi_Book}. The individual edge loop HLLs are described by
\begin{align}
	\hat{H}_{b,0}
	=
	\frac{\hbar v}{2}\sum_{a=1,2}\int dx\left[K\left(\partial_x\phi_a\right)^2+\frac{1}{K}\left(\partial_x\theta_a\right)^2\right],
\end{align}
where $K$ is the Luttinger parameter and $v$ is the 
charge velocity. $K=1$ and $v = v_F$ corresponds to the free fermion limit. The density ($n$) and current ($I$) can be expressed in terms of the axial fields as
$	n_a = \partial_x\theta_a / \sqrt{\pi}$ and
$	I_a= - \partial_{t}\theta_a / \sqrt{\pi}$, respectively. 
The interedge interaction $\hat{H}_-$ is bosonized to
\begin{align}
	\label{H_b_-}
	\hat{H}_{b,-}
	=
	\frac{U_-}{2 \pi^{2}\alpha^2}\int dx\cos\left[\sqrt{4\pi}\left(\theta_1-\theta_2\right)\right],
\end{align}
where $\alpha$ is an ultraviolet length scale.


{\it Perfect current drag and dc conductance}. -- 
At zero temperature, two infinite HLLs
form an interedge locking state \cite{Nazarov1998,Klesse2000} due to the two-particle backscattering term in Eq.~(\ref{H_b_-}). 
The locking state is characterized by 
	$
		\theta_1(t,x)=\theta_2(t,x) + c_m,
	$ 
where
	$	
		c_m = (m + 1/2)\sqrt{\pi}
	$ 
is a constant and $m \in \mathbb{Z}$. This state exhibits perfect current drag \cite{Nazarov1998}, 
$I_1 = I_2$ in Fig.~\ref{Fig:2Edge_scheme}. The conductance of the locked edges (both carrying current $I_1$) is $I_1/V = (e^2/h)[K / (K + 1)]$. This can be understood as the series resistor combination of a 
spinless LL connected to leads with resistance $h/e^2$ \cite{Maslov1995,Ponomarenko1995,Safi1995}
and one with periodic boundary conditions and resistance $h / K e^2$ \cite{ApelRice82,KaneFisher92}. 
An explicit Green's function calculation confirms this result \cite{SUPINFO}, which is also independent of disorder. Equation (\ref{GResult}) is obtained by adding the parallel $I_1'$ edge channel.

For a finite interacting region of length $L$ and nonzero temperature $T$, we require that $L \gg \xi$ and $k_B T \ll \Delta$. Occasional phase slips between the drive and slave circuits give rise to corrections that are exponentially small in $L / \xi$ and $\Delta / k_B T$ \cite{Nazarov1998,Ponomarenko2000,Klesse2000}. 
For $L = 1 \text{ $\mu$m}$ in InAs/GaSb with  $v \sim v_F = 3 \times 10^4 \text{ m/s}$, this gives a lower bound for $\Delta$ of the order of $\hbar v / L = 0.02 \text{ meV}$. We assume that $k_B T$ is larger than the latter to avoid coherent instanton effects \cite{Ponomarenko2000}. By comparison, the bulk minigap 
is of the order of $4 \text{ meV}$ \cite{Knez2011}. The Mott gap takes the form \cite{Giamarchi_Book}
$
	\Delta \sim \sqrt{K \, U_{-} \hbar v} \, / \alpha.
$
Using $\alpha = 1 \text{ nm}$ gives $\Delta \sim 20 \text{ meV} \, \sqrt{K \, (U_{-} / \hbar v)}$.
The interaction strength $U_-$ is obtained from the inter-edge Coulomb potential,
mediated by matrix elements determined by the Rashba SOC in each TI (since it vanishes in its absence).
The result will depend on microscopic details that we do not analyze here.


{\it Finite temperature corrections}. -- Above a crossover temperature  $T^* \sim \Delta/k_B$\cite{Klesse2000}, inelastic electron-electron collisions due to the interedge interactions 
in Eqs.~(\ref{H_I_-})--(\ref{H_U}) can be treated perturbatively. In addition, we consider intraedge collisions due to electron-electron interactions [Eq.~(\ref{H_Ine}), below] and forward-scattering potential disorder.
Ordinary backscattering (random mass) disorder is forbidden by time-reversal symmetry.
We ignore irrelevant backscattering disorder terms with extra derivatives that are not expected to impact
the conductivity in isolation \cite{Kainaris2014} and which give subleading corrections in combination
with interactions. Forward-scattering disorder is encoded in 
	$
		\hat{H}_{\text{imp}}=\sum_{a=1,2}\int dx\, \eta_a(x) \, n_a(x),
	$
where $\eta_a(x)$ is a random potential obeying $\overline{\eta_a(x)}=0$ and $\overline{\eta_a(x)\eta_b(x')} = g_{\eta} \, \delta_{a,b}\delta(x-x')$. The $\overline{\cdots}$ denotes disorder averaging, while $g_{\eta}$ characterizes the disorder strength.

To compute the conductivity, we evaluate interaction corrections to the inverse boson propagator via the 
effective potential method \cite{Peskin_Book,Ristivojevic12}. We use replicas to average over disorder. 
The retarded boson correlation function is
\begin{align}\label{Dyson_2edge}
	\left[\hat{\mathcal{G}}^{(R)}(\omega,k)\right]_{ab}^{-1}
	=
	\left[\hat{G}^{(R)}(\omega,k)\right]^{-1}_{ab}
	-
	\left[\hat{\Pi}^{(R)}(\omega,k)\right]_{ab},
\end{align}
where $a,b \in \{1,2\}$ indicate the edges. The noninteracting propagator is $\hat{G}^{(R)}(\omega,k)$,
while $\hat{\Pi}^{(R)}$ denotes the self-energy describing the interaction corrections. Equation (\ref{Dyson_2edge}) is a matrix Dyson equation. At second order in the coupling constants, $\hat{\Pi}^{(R)}$ contains an imaginary part that determines the scattering rates; the real part does not 
contribute to dc conductivity. In the limit $\omega \rightarrow 0$ with $k = 0$,
\begin{align}\label{Pi_2_Xi}
	\text{Im}\left[\Pi_{ab}^{(R)}\right]
	=
	-2\omega\Xi_{ab}
	+
	\ord{\omega^2},
\end{align}
where $\Xi_{ab}$ is the ``rate'' (inverse mean free path) associated to $\Pi_{ab}$. The components are 
	$
		\Xi_{11}=\Xi_{22}=\frac{1}{2}\Xi_++\frac{1}{2}\Xi_-+\Xi_U+\Xi_{W}
	$ 
	and 
	$
		\Xi_{12}=\Xi_{21}=\frac{1}{2}\Xi_+-\frac{1}{2}\Xi_-
	$.  
$\Xi_U$ is due to the one-particle backscattering term $\hat{H}_U$. $\Xi_{+}$ and $\Xi_{-}$ correspond to the two-particle backscattering interactions $\hat{H}_{+}$ and $\hat{H}_{-}$, respectively. $\Xi_{W}$ is due to intraedge inelastic electron-electron collisions, $\hat{H}_{W}$ in Eq.~(\ref{H_Ine}). $\Xi_U$ and $\Xi_{W}$ affect only the diagonal elements of the self-energy, while $\Xi_+$ and $\Xi_-$ contribute to all of the components.  

Temperature dependences of all the scattering rates are obtained analytically. For
$k_B T \gg \max(g_{\eta} / \hbar v, \hbar v k_F)$, 
$\Xi_U \propto T^{2K-1}$, 
$\Xi_{W}\propto T^{2K+1}$, 
and 
$\Xi_{\pm}\propto T^{4K-3}$.
In the presence of disorder
and for 
$k_B T  \ll g_{\eta} / \hbar v$, 
$\Xi_U \propto T^{2K}$, 
$\Xi_{W}\propto T^{2K+2}$, 
and 
$\Xi_{\pm}\propto T^{4K-2}$. The additional power of $T$ comes
from disorder smearing of $k_F$ \cite{Fiete2006}.
Full crossovers with and without disorder are determined by the explicit forms 
of $\Xi_{U,W,\pm}$ provided in Ref.~\cite{SUPINFO}.

The dc conductivity can be obtained through Kubo formula \cite{Sirker2011}. The intraedge dc conductivity is
\begin{align}
	\nonumber
	\sigma_{11}
	&=
	-\frac{1}{\pi}\frac{e^2}{\hbar}\lim_{\omega\rightarrow 0}\text{Im}\left[\omega\,\mathcal{G}^{(R)}_{11}(\omega,k)\right]
	\\
	\label{dc_cond_2edge}
	&=
	\frac{e^2}{2h}\left[\frac{1}{\Xi_++\Xi_U+\Xi_{W}}+\frac{1}{\Xi_-+\Xi_U+\Xi_{W}}\right].
\end{align}
This expression is very different from the conductivity of an isolated edge, discussed below.
Both intraedge and interedge interactions contribute to Eq.~(\ref{dc_cond_2edge}),
but the intraedge contribution $\Xi_{W}$ is subleading comparing to 
the interedge rates. The finite-temperature behavior for $\sigma_{11}$ is summarized as follows. 
For clean edges and $k_B T \gg \hbar v k_F$,
\begin{align}
\sigma_{11}\sim\begin{cases}
T^{-4K+3}, & \text{for } K\le 1,\\
T^{-2K+1}, & \text{for } K>1.
\end{cases}
\end{align}
With smooth disorder and $k_B T \ll g_{\eta} / \hbar v$,
\begin{align}
\sigma_{11}\sim\begin{cases}
T^{-4K+2}, & \text{for } K\le 1,\\
T^{-2K}, & \text{for } K>1.
\end{cases}
\end{align}
$K\le 1$ (repulsive interactions) is the physical situation.


The transconductivity is
\begin{align}\label{drag_cond}
	\sigma_{21}=\frac{e^2}{2h}\left[\frac{1}{\Xi_++\Xi_U+\Xi_{W}}-\frac{1}{\Xi_-+\Xi_U+\Xi_{W}}\right].
\end{align}
This can be measured by shorting the distal part of passive edge loop with an ideal (zero input impedance) current meter. The leading temperature dependence of the drag conductivity is the same as the intraedge conductivity. In the usual case, one measures instead the drag resistivity \cite{Rojo1999,Narozhny2015}.
Here this evaluates to 
$
	\rho_{D}=-\rho_{12}=(h/2 e^2)\left[\Xi_- - \Xi_+\right],
$
independent of the interedge $U$ and intraedge $W$ interactions. In the case of clean identical edges, a positive drag resistivity with leading $T^{4K-3}$ behavior is obtained. This is the same result found previously for spinless Luttinger liquids \cite{Klesse2000,Ponomarenko2000} and TI edges with small magnetic fields \cite{Zyuzin10}.


{\it Single edge}. -- Finally, we consider dc conductivity of a single edge in isolation, in the presence of Rashba SOC. The \emph{least irrelevant} intra-edge electron-electron interaction term allowed by time-reversal symmetry that can give a finite transport lifetime
is 
\begin{align}
	\nonumber
	\hat{H}_{W}=&\,
	W
	\int dx\,
	:
	\big\{
		e^{i2k_Fx}L^{\dagger}(x)R(x)R^{\dagger}(x)[-i\partial_xR(x)]\\
	\label{H_Ine}
		+
		&
		e^{-i2k_Fx}R^{\dagger}(x)L(x)L^{\dagger}(x)[-i\partial_xL(x)]
		+
		\text{H.c.}
	\big\}\! :,\!\!
\end{align}
where $:\!\!\mathcal{O}\!\!:$ denotes the normal ordering of $\mathcal{O}$. This is a one-particle spin-flip umklapp term. Similar one-particle backscattering interactions appear in Refs.~\cite{Schmidt2012,Lezmy2012,Kainaris2014}, but the full temperature dependence of the dc conductivity was not determined. The interaction correction due to Eq.~(\ref{H_Ine}) can be described by a self-energy with imaginary part
	$
		\text{Im}\left[\Pi_{W}^{(R)}(\omega,k)\right]
		= 
		-
		2
		\omega\Xi_{W}
		+
		\ord{\omega^2},
	$ 
when $k = 0$ and $\omega\rightarrow 0$. We find that
\begin{align}
	\nonumber
	\Xi_{W}
	=&
	\frac{\tilde{W}^2 \, \alpha^{2K}}{(\hbar v)^2 \, l_T^{2 K + 1}}
	\frac{2^{2K}\pi^{2K+3} K\Gamma\left[-K-3\right]}{\Gamma\left[K+2\right]}\\
\nonumber&
	\times
	\int_{-\infty}^{\infty} d y
		\left[
			\frac{\gamma/\pi}{\left(y - \frac{k_F l_T}{2\pi}\right)^2+\gamma^2}
			+
			(k_F \rightarrow - k_F)
		\right]\\
\label{Xi_Delta}&
	\times
	\frac{\sin\left(\pi K\right)}{\cosh\left(2\pi y\right)-\cos\left(\pi K\right)}
	\left|\frac{\Gamma\left[\frac{4+K}{2}+i y\right]}{\Gamma\left[\frac{2-K}{2}+i y\right]}\right|^2,
\end{align}
where $\tilde{W}=W/(\pi^{3/2}\alpha)$ and $l_T \equiv \hbar v / k_B T$ denotes the thermal de Broglie wavelength. The disorder is encoded in $\gamma \equiv l_T (K / \hbar v)^2 g_{\eta}/2\pi$. The  dc conductivity is 
$
	\sigma_{\text{dc}}=(e^2/h)(1/\Xi_{W}).
$
At zero temperature where the HLL exhibits ballistic transport, $\sigma_{\text{dc}}$  diverges.

For a clean noninteracting edge ($K=1$ and $v = v_F$), the conductivity reduces to 
\begin{align}\label{dc_cond_free}
	\sigma_{\text{dc}}=\frac{e^2}{h}\frac{(\hbar v)^2 \, l_T^3}{W^2 \pi^3}
	\frac{
		6\left[\cosh\left(k_F l_T\right)+1\right]
		}{
	\left[	(\frac{k_F l_T}{2\pi})^4 + \frac{5}{2}(\frac{k_F l_T}{2\pi})^2 + \frac{9}{16}	\right]
	}.
\end{align}
At high temperatures $k_F l_T \gg 1$, this is proportional to $T^{-3}$;
in the opposite limit the umklapp scattering is thermally activated, giving
$\sigma_{\text{dc}} \sim T\exp\left(k_F l_T\right)$. 
Luttinger interactions modify the high-temperature behavior to $T^{-2K-1}$,
while disorder leads to $\sigma_{\text{dc}} \sim T^{-2K-2}$ for 
$l_T \gg (\hbar v)^{2} g_{\eta}^{-1}$, again due to smearing of the Fermi 
momentum \cite{Fiete2006}. The disordered result is consistent with earlier predictions \cite{Schmidt2012,Lezmy2012,Kainaris2014}. The $T^{-2K-1}$ behavior is the most important temperature dependence in the clean edge due to the intraedge interactions, in the presence of Rashba SOC. 
The responsible interaction term in Eq.~(\ref{H_Ine}) will be generated by renormalization irrespective of whether it arises in a particular microscopic model.


{\it Summary and discussion} - In this work, we have shown that low-temperature edge state transport measurements for two proximate HLLs can quantify the value of the Luttinger parameter in the presence of spin-flip interedge electron-electron scattering. The latter is enabled by Rashba SOC within each TI, as can arise in InAs/GaSb. In contrast to the usual setup for Coulomb drag, the passive circuit floats without leads and provides a much stronger source of scattering for the active circuit edge than intraedge interactions, which are negligible at low temperature. Because of the topological protection, this result is immune to disorder but receives exponentially small corrections for a long, but finite interacting region.  

In the same device geometry, both the intraedge conductivity and the transconductivity show the same leading temperature dependence for $T$ above the crossover scale to the low-temperature locking regime.
Thus, two-terminal conductivity gives an alternative route to detect Coulomb drag physics. We have
also computed the conductivity correction due to the least irrelevant symmetry-allowed interaction in a given edge. This gives $T^{-2K-1}$ temperature dependence for a clean edge.

We close with some observations and avenues for future work. In general, negative drag is possible when $|k_{F1}-k_{F2}| \gg |k_{F1}+k_{F2}|$. Eq.~(\ref{H_I_+}) instead of Equation (\ref{H_I_-}) dominates the interedge interactions at low temperature in this case. For two almost identical edges but $k_{F1}=-k_{F2}$, a perfect antiparallel current locking can occur; the two-terminal conductance is still given by Eq.~(\ref{GResult}) in the $T \rightarrow 0$ limit. The finite-temperature behavior will be qualitative the same as the parallel drag situation. The generic kinetic theory of Coulomb drag between helical edge states, that also includes the forward-scattering long-ranged component of the Coulomb interaction, is an important topic for future work \cite{Niko}. Understanding how a HLL edge state thermalizes via the various scattering mechanisms has crucial implications for nonequilibrium spectroscopy \cite{Altimiras2010_NP,Apostolov}. It will also be interesting to study the noise \cite{Blanter2000} for the two-helical-edge setup
described here. 

Y.-Z.C. and M.S.F. thank R.-R. Du, L. Du, D. Natelson, and A. Nevidomskyy for useful discussions. A.L. thanks N. Kainaris, I. Gornyi and D. Polyakov for multiple important discussions and ongoing collaboration on a related problem. A.L. and M.S.F. acknowledge the hospitality of the Spin Phenomena Interdisciplinary Center (SPICE), where this work was completed. Y.-Z.C. and M.S.F. acknowledge funding from the Welch Foundation under Grant No. C-1809 and from an Alfred P. Sloan Research Fellowship (No. BR2014-035). Y.-Z.C. also acknowledges hospitality of the Michigan State University. A.L. acknowledges funding from NSF Grants No. DMR-1401908 and No. ECCS-1407875.



\newpage \clearpage 

\onecolumngrid

\begin{center}
	{\large
	Helical Quantum Edge Gears in 2D Topological Insulators:
	\vspace{4pt}
	\\
	SUPPLEMENTAL MATERIAL
	}
\end{center}

We set $\hbar = k_B = 1$ except as noted. 

\section{Bosonization Conventions}

We adopt the standard field theoretic bosonization method. The fermionic fields can be described by chiral bosons via
\begin{align}
R(x)=\frac{1}{\sqrt{2\pi\alpha}}e^{i\sqrt{\pi}\left[\phi(x)+\theta(x)\right]},\,\,L(x)=\frac{1}{\sqrt{2\pi\alpha}}e^{i\sqrt{\pi}\left[\phi(x)-\theta(x)\right]},
\end{align}
where $\alpha$ is some ultraviolet length scale.

The time reversal operations in the bosonic language read: $\phi\rightarrow -\phi+\frac{\sqrt{\pi}}{2}$, $\theta\rightarrow\theta-\frac{\sqrt{\pi}}{2}$, and $i\rightarrow-i$.

\section{Bosonic Actions}

In the imaginary time formalism, the Luttinger action for the $a$th edge reads
\begin{align}
	\mathcal{S}_{0,a}
	=&
	\int\limits_{\tau,x}i\left(\partial_x\theta_a\right)
	\left(\partial_{\tau}\phi_a\right)+\frac{v}{2}\int\limits_{\tau,x}\left[K\left(\partial_x\phi_a\right)^2+\frac{1}{K}\left(\partial_x\theta_a\right)^2	\right],
	\\
	\mathcal{S}_{W,a}
	=&
	\tilde{W}_a\int\limits_{\tau,x}\left(\partial_x^2\phi_a\right)
	\cos\left(\sqrt{4\pi}\,\theta_a+2k_{F,a}x\right),
	\\
	\mathcal{S}_{\text{imp},a}
	=&
	\int\limits_{\tau,x}\,\eta_a(x)\frac{1}{\sqrt{\pi}}\partial_x\theta_a,
\end{align}
where $\int\limits_{\tau,x}$ is the short hand notation for $\int d\tau dx$.

The inter-edge interactions correspond to
\begin{align}	
	\mathcal{S}_{U}
	&=
	\sum_{a=1,2}
	\frac{U_a}{\pi^{3/2}\alpha}
	\int\limits_{\tau,x}
	\left(\partial_x\phi_{\bar{a}}\right)\sin\left[\sqrt{4\pi}\theta_a+2k_{Fa}x\right],
\\
	\mathcal{S}_{+}
	&=
	\frac{-U_+}{2 \pi^2 \alpha^2}
	\int\limits_{\tau,x}
	\cos\left[\sqrt{4\pi}\left(\theta_1+\theta_2\right)+2(k_{F1}+k_{F2})x\right],
\\
	\mathcal{S}_{-}
	&=
	\frac{U_-}{2 \pi^2 \alpha^2}
	\int\limits_{\tau,x}
	\cos\left[\sqrt{4\pi}\left(\theta_1-\theta_2\right)+2(k_{F1}-k_{F2})x\right],
\end{align}
where $a$ labels the edge, with $\bar{1} = 2$ and $\bar{2} = 1$.

We can integrate over $\partial_x \phi_1$ and $\partial_x \phi_2$ exactly. The $\theta$-only action is
\begin{align}
	\mathcal{S}_{\theta}
	=&\,
	\sum_{a=1,2}\frac{1}{2v_a K_a}
	\int\limits_{\tau,x}
	\left[\left(\partial_{\tau}\theta_a\right)^2+v_a^2\left(\partial_{x}\theta_a\right)^2\right]
\\
	&-
	\sum_{a=1,2}\frac{i\tilde{W}_a}{v_aK_a}
	\int\limits_{\tau,x}
	\left(\partial_{\tau} \partial_x \theta_a\right) 
	\, 
	\cos\left(\sqrt{4\pi}\,\theta_a+2k_{Fa}x\right)
\\
	&-
	\sum_{a=1,2}\frac{iU_{\bar{a}}}{v_aK_a\pi^{3/2}\alpha }
	\int\limits_{\tau,x}
	\left(\partial_{\tau}\theta_a\right)\sin\left(\sqrt{4\pi}\theta_{\bar{a}}+2k_{F\bar{a}}x\right)
	+ 
	\mathcal{S}_{+}
	+ 
	\mathcal{S}_{-},
\end{align}
where we have dropped terms not contributing to the boson self-energy at the second homogeneous order of the coupling constants. 

Forward-scattering potential disorder is averaged using the replica trick. 
The quadratic part of the action becomes
\begin{align}
	\mathcal{S}_{\theta,0}
	=
	\frac{1}{2v K}
	\sum_{a=1,2}
	\sum_{n=1}^R
	\int\limits d\tau d x
	\left[\left(\partial_{\tau}\theta_{a,n}\right)^2+v^2\left(\partial_{x}\theta_{a,n}\right)^2\right]
	-
	\frac{g_{\eta}}{2\pi}
	\sum_{a=1,2}
	\sum_{n,m=1}^R
	\int\limits d\tau d\tau' d x 
	\left[\partial_x\theta_{a,n}(\tau)\right]
	\left[\partial_x\theta_{a,m}(\tau')\right],
\end{align}
where $n$, $m$ are the replica indexes, $R$ is the number of replica, and $g_{\eta}$ is the variance of 
the disorder potential [$\eta(x)$, introduced in the main text]. All interaction terms are diagonal in the replica space.

\section{dc Conductance}

We derive the zero temperature dc conductance result [Eq.~(1)] in this section. We consider the device geometry in Fig.~1. The edge carrying current $I_1$ has Luttinger parameter $K$ and charge velocity $v$.
It is connected to external leads, which we treat as non-interacting free fermion reservoirs with $K = 1$ and $v = v_F$. The electric field is applied between the leads. The passive circuit edge carrying current $I_2$ forms a closed loop with uniform Luttinger parameter $K$ and velocity $v$. For simplicity, we assume that this loop is infinitely long (but see below). The two edges interact via Eq.~(7) in the region $-L/2\le x\le L/2$. 
At zero temperature, the locking condition holds across this span of length $L$. We can replace the sine-Gordon term in Eq.~(7) by a mass term,
\begin{align}
	\hat{H}_{b,M}
	=
	\frac{M^2}{2}
	\int_{-L/2}^{L/2}
	d x
	\,
	\left[\frac{\theta_1(x)-\theta_2(x)-c_0}{\sqrt{2}}\right]^2.
\end{align}
The constant $c_0$ can be absorbed by shifting $\theta_2\rightarrow\theta_2-c_0$;
the locking condition becomes $\theta_1=\theta_2$. 

The current $I_1$ can be expressed in terms of retarded Green's functions,
\begin{align}
	\label{current}
	\langle I_1(x)\rangle
	=
	i\frac{e^2}{\pi\hbar}
	\int_{-L/2}^{L/2} d x'
	\left[E-\partial_{x'}\eta_1(x')\right]
	\lim\limits_{\omega\rightarrow 0}
	\left[\omega \,\tilde{G}_{11}^{(R)}(\omega;x,x')\right]
	-
	i\frac{e^2}{\pi\hbar}
	\int_{-\infty}^{\infty} d x'
	\lim\limits_{\omega\rightarrow 0}
	\left[\partial_{x'}\eta_2(x')\right]
	\left[\omega \,\tilde{G}_{12}^{(R)}(\omega;x,x')\right],
\end{align}
where $E$ is the external electric field in the region $-L/2\le x\le L/2$, and $\eta_{1,2}(x)$ are the random forward scattering potentials. The retarded Green's functions in the above formula are determined by
\begin{align}
	\left[
	\begin{array}{cc}
		\frac{\omega^2}{v_1(x)K_1(x)}+\partial_x\left[\frac{v_1(x)}{K_1(x)}\partial_{x}\right]-\frac{M^2(x)}{2}& \frac{M^2(x)}{2} \\[2mm]
		\frac{M^2(x)}{2} & \frac{\omega^2}{v K}+\frac{v}{K}\partial_{x}^2-\frac{M^2(x)}{2}
	\end{array}
	\right]
	\left[
	\begin{array}{cc}
		\tilde{G}^{(R)}_{11}(\omega;x,x') & \tilde{G}^{(R)}_{12}(\omega;x,x')\\[2mm]
		\tilde{G}^{(R)}_{21}(\omega;x,x') & \tilde{G}^{(R)}_{22}(\omega;x,x')
	\end{array}
	\right]=\delta(x-x')\hat{1},
\end{align}
where
\begin{align}
K_1(x)&=\begin{cases}
K, & -L/2\le x\le L/2\\
1, & |x|>L/2
\end{cases}\\
v_1(x)&=\begin{cases}
v, & -L/2\le x\le L/2\\
v_F, & |x|>L/2
\end{cases}\\
M(x)&=\begin{cases}
M, & -L/2\le x\le L/2\\
0, & |x|>L/2
\end{cases}
\end{align}

The retarded Green function can be solved by imposing the following boundary conditions \cite{Maslov1995,Maslov}.
	(i) 	$\tilde G_{ab}^{(R)}$ is continuous everywhere in $x$. 
	(ii) 	$\frac{v_1(x)}{K_1(x)}\partial_x\tilde G_{11}^{(R)}$, 
		$\frac{v_1(x)}{K_1(x)}\partial_x\tilde G_{12}^{(R)}$, 
		$\partial_x\tilde G_{21}^{(R)}$, and 
		$\partial_x\tilde G_{22}^{(R)}$ are continuous for $x\neq x'$. 
	(iii) 	For $x \sim x'$,
\begin{align}
&\left[\frac{v_1(x)}{K_1(x)}\partial_x\tilde G_{11}^{(R)}(\omega;x,x')\right]_{x=x'+0^+}-\left[\frac{v_1(x)}{K_1(x)}\partial_x\tilde G_{11}^{(R)}(\omega;x,x')\right]_{x=x'-0^+}=1,\\
&\left[\frac{v_1(x)}{K_1(x)}\partial_x\tilde G_{12}^{(R)}(\omega;x,x')\right]_{x=x'+0^+}-\left[\frac{v_1(x)}{K_1(x)}\partial_x\tilde G_{12}^{(R)}(\omega;x,x')\right]_{x=x'-0^+}=0,\\
&\left[\frac{v}{K}\partial_x\tilde G_{21}^{(R)}(\omega;x,x')\right]_{x=x'+0^+}-\left[\frac{v}{K}\partial_x\tilde G_{21}^{(R)}(\omega;x,x')\right]_{x=x'-0^+}=0\\
&\left[\frac{v}{K}\partial_x\tilde G_{22}^{(R)}(\omega;x,x')\right]_{x=x'+0^+}-\left[\frac{v}{K}\partial_x\tilde G_{22}^{(R)}(\omega;x,x')\right]_{x=x'-0^+}=1.
\end{align}
(iv) The retarded Green functions obey the outgoing wave conditions.\\

Expanding $\tilde G_{ab}^{(R)}$ in each region in terms of propagating and/or evanescent waves and imposing 
all boundary conditions, we find that
\begin{align}
\lim\limits_{\omega\rightarrow 0}\left[\omega \,\tilde{G}_{11}^{(R)}(\omega;x,x')\right]=\lim\limits_{\omega\rightarrow 0}\left[\omega \,\tilde{G}_{12}^{(R)}(\omega;x,x')\right]=-i\frac{K}{2(1+K)}.
\end{align}
The result is independent of $x$. The current expression in Eq.~(\ref{current}) becomes
\begin{align}
\nonumber\langle I_1(x)\rangle&=\frac{e^2}{\pi\hbar}\frac{K}{2(1+K)}\int_{-L/2}^{L/2}dx'\left[E-\partial_{x'}\eta_1(x')\right]-\frac{e^2}{\pi\hbar}\frac{K}{2(1+K)}\int_{-\infty}^{\infty}dx'\left[\partial_{x'}\eta_2(x')\right]\\
&=\frac{K}{(1+K)}\frac{e^2}{h}\left[EL-\eta_1(L/2)+\eta_1(-L/2)\right]-\frac{K}{(1+K)}\frac{e^2}{h}\left[\eta_2(\infty)-\eta_2(-\infty)\right],
\end{align}
where $\eta_1(L/2)=\eta_1(-L/2)=0$ because the first edge is connected to free fermion leads, 
and  $\eta_2(\infty) = \eta_2(-\infty)$ due to the periodic boundary condition. The conductance of the locked edges is then
\begin{align}
	G=\frac{\langle I_1\rangle}{EL}=\frac{K}{1+K}\frac{e^2}{h}.
\end{align}
Adding the contribution of the parallel edge carrying $I_1'$ gives Eq.~(1).

In the presence of the inter-edge Luttinger interactions, the Luttinger parameter in the region 
$-L/2\le x\le L/2$ is modified. The two-terminal conductance is unchanged if we assume that the distal part of the passive edge is much longer than the interacting part, as above. Then this passive edge is effectively connected to the Luttinger liquid leads with Luttinger parameter $K$, and this gives resistance $(1/K)h/e^2$. 
For a finite total length $l > L$ of the passive edge, the value of the measured Luttinger parameter is between the intra-edge value $K$ and the Luttinger parameter for the symmetric mode, and it also depends on the ratio of the interacting region length $L$ to the total length $l$.

\section{Scattering Rates}

At second order in the interaction coupling strengths, there are four distinct self-energies, 
$\Pi_{W}$, $\Pi_U$, $\Pi_+$, and $\Pi_-$. Coupling constant mixing will occur at higher orders, but
is prevented here by vertex operator charge neutrality conditions. We are interested in the imaginary part of the retarded self-energies. In the long-wavelength and low-energy limits, $\text{Im}\left[\Pi^{(R)}(\omega,k)\right]\approx-2\omega\Xi$.

In this section we provide the explicit results for the scattering rates that enter the intra-edge and transconductivities in Eqs.~(10) and (13). The scattering rate due to Eq.~(14) already appears in Eq.~(15). Here $\gamma = K^2\beta g_{\eta} / 2\pi v$ is a parameter indicating the ratio of the effective disorder strength to the temperature.

In the clean limit ($\gamma\rightarrow 0$), the Lorentzian distributions in Eq.~(15) becomes delta functions. 
The scattering rate reduces to
\begin{align}
\Xi_{W}=\left(\tilde{W}\alpha^{K}\right)^2\frac{2^{2K+1}\pi^{2K+3}}{v^{2K+3}\beta^{2K+1}} K\frac{\Gamma\left[-K-3\right]}{\Gamma\left[K+2\right]}\frac{\sin\left(\pi K\right)}{\cosh\left(v\beta k_{F}\right)-\cos\left(\pi K\right)}\left|\frac{\Gamma\left[\frac{4+K}{2}+i\frac{v\beta k_{F}}{2\pi}\right]}{\Gamma\left[\frac{2-K}{2}+i\frac{v\beta k_{F}}{2\pi}\right]}\right|^2,
\end{align}
where $\beta$ is the inverse temperature.
When $T \gg v k_{F}$, the clean scattering rate is proportional to $T^{2K+1}$.

In the presence of disorder and at low temperatures such that $\gamma\gg 1$, 
the scattering rate becomes
\begin{align}
	\Xi_{W}
	=
	\left(\tilde{W}\alpha^K\right)^2
	\frac{2^{2K}\pi^{2K+1}}{v^{2K+3}\beta^{2K+2}}
	\left[	
		\frac{\beta\gamma/\pi}{\left(\frac{v\beta k_{F}}{2\pi}\right)^2+\gamma^2}	
	\right]
	K\sin\left(\pi K\right)\frac{\sqrt{\pi}\,\Gamma\left[2+\frac{K}{2}\right]\Gamma\left[\frac{5}{2}+\frac{K}{2}\right]
	\Gamma\left[\frac{K}{2}\right]\Gamma\left[\frac{1}{2}+\frac{K}{2}\right]\Gamma\left[-K-3\right]}{\Gamma\left[\frac{5}{2}+K\right]}.
\end{align}
The term in square brackets is independent of temperature, so that $\Xi_{W} \sim T^{2K+2}$.

The inter-edge single particle backscattering Hamiltonian in Eq.~(5) contains two terms, $U_1$ and $U_2$. 
$U_1$ and $U_2$ correct the dc conductivity in edge 1 and 2, respectively. The scattering rates are
\begin{align}
\nonumber\Xi_{U,a}=&\left(\tilde{U}_a\alpha^{K}\right)^2\frac{2^{2K-3}\pi^{2K+1}}{v^{2K+1}\beta^{2K-1}}
\frac{1}{K}\frac{\Gamma\left[1-K\right]}{\Gamma\left[2+K\right]}\\
&\times\int_{-\infty}^{\infty}  dy \left\{
\left[\frac{\gamma/\pi}{\left(y-\frac{v\beta k_{F,a}}{2\pi}\right)^2+\gamma^2}
+\frac{\gamma/\pi}{\left(y+\frac{v\beta k_{F,a}}{2\pi}\right)^2+\gamma^2}\right]\frac{\sin\left(\pi K\right)}
{\cosh\left(2\pi y\right)-\cos\left(\pi K\right)}
\left|\frac{\Gamma\left[\frac{K}{2}+iy\right]}
{\Gamma\left[-\frac{K}{2}+iy\right]}\right|^2
\right\},
\end{align}
where 
$\tilde{U}_a = U_a / \pi^{3/2}\alpha$. 
In the high-temperature limit $\gamma\rightarrow 0$ and $v\beta k_{F,a}\rightarrow 0$, 
$\Xi_U$ is proportional to $T^{2K-1}$. 
In the low-temperature limit $\gamma \gg 1$, 
$\Xi_U$ is proportional to $T^{2K}$. 
In the main text, we assume two identical edges and $U_1=U_2$, so that $\Xi_{U,1}=\Xi_{U,2}=\Xi_{U}$.

The scattering rates due to Eqs.~(3) and (4) are given by
\begin{align}
	\nonumber
	\Xi_{\pm}
	=&
	\left(\tilde{U}_{\pm} \alpha^{2K}\right)^2
	\frac{2^{4K-3}\pi^{4K}}{v^{4K-1}\beta^{4K-3}}
	\frac{\Gamma\left[1-2K\right]}{\Gamma[2K]}
	\\
	&\times\int_{-\infty}^{\infty} dy
	\left\{
	\left[
		\frac{\bar{\gamma}/\pi}{\left(y-\frac{v\beta k_{F}^{\pm}}{4\pi}\right)^2+\bar{\gamma}^2}
		+
		\frac{\bar{\gamma}/\pi}{\left(y+\frac{v\beta k_{F}^{\pm}}{4\pi}\right)^2+\bar{\gamma}^2}
	\right]
	\frac{\sin(2 \pi K) }{\cosh\left(2\pi y\right)-\cos(2\pi K)}
	\left|\frac{\Gamma \left[K+iy\right]}{\Gamma \left[1-K+iy\right]}\right|^2
	\right\},
\end{align}
where $\bar{\gamma}=K^2\beta g_{\eta}/\pi v$, $\tilde{U}_{\pm} = \mp U_{\pm} / 2 \pi^2 \alpha^2$, 
and $k_{F}^{\pm}=2(k_{F1} \pm k_{F2})$. In the high-temperature limit $\bar\gamma\rightarrow 0$ and $v\beta k_{F}^{\pm} \rightarrow 0$, $\Xi_{\pm}$ is proportional to $T^{4K-3}$. In the low-temperature limit $\bar\gamma \gg 1$, it is proportional to $T^{4K-2}$.

\end{document}